\documentclass[11pt]{article}

\usepackage[utf8]{inputenc}
\usepackage[T1]{fontenc}
\usepackage{lmodern}
\usepackage[margin=1in]{geometry}
\usepackage[hyphens,spaces,obeyspaces]{url}
\usepackage[breaklinks]{hyperref}
\usepackage{booktabs}
\usepackage{graphicx}
\usepackage{listings}
\usepackage{xcolor}
\usepackage{amsmath,amssymb}
\usepackage{enumitem}
\usepackage{caption}
\usepackage{balance}
\usepackage{tikz}
\usetikzlibrary{arrows.meta, positioning, calc}

\newcommand{\tool}[1]{\texttt{#1}}
\newcommand{\layer}[1]{\textbf{L#1}}
\newcommand{\rw}{\textsc{Resilient Write}}

\title{Resilient Write: A Six-Layer Durable Write Surface\\for LLM Coding Agents}
\author{%
Justice Owusu Agyemang$^{1,2,3}$\thanks{\texttt{jay@sperixlabs.org, jay@knust.edu.gh}},\quad
Jerry John Kponyo$^{2}$\thanks{\texttt{jjkponyo.soe@knust.edu.gh}},\quad
Elliot Amponsah$^{2}$\thanks{\texttt{eamponsah52@st.knust.edu.gh}},\\
Godfred Manu Addo Boakye$^{2}$\thanks{\texttt{gmaboakye@st.knust.edu.gh}},\quad
Kwame Opuni-Boachie Obour Agyekum$^{3}$\thanks{\texttt{kooagyekum@knust.edu.gh}}\\[8pt]
\small
$^1$Sperixlabs, Ghana\\
$^2$Kwame Nkrumah University of Science and Technology, Kumasi, Ghana\\
$^3$VIA Cybersecurity Lab, Kwame Nkrumah University of Science and Technology, Kumasi, Ghana
}
\date{June 2026}

\begin{document}
\maketitle

\begin{abstract}
LLM-powered coding agents increasingly rely on tool-use protocols
such as the Model Context Protocol~(MCP) to read and write files
on a developer's workstation.  When a write fails---due to content
filters, truncation, or an interrupted session---the agent typically
receives no structured signal, loses the draft, and wastes tokens
retrying blindly.  We present \textbf{Resilient Write}, an MCP
server that interposes a six-layer durable write surface between the
agent and the filesystem.  The layers---pre-flight risk scoring,
transactional atomic writes, resume-safe chunking, structured typed
errors, out-of-band scratchpad storage, and task-continuity
handoff envelopes---are orthogonal and independently adoptable.
Each layer maps to a concrete failure mode observed during a real
agent session in April~2026, in which content-safety filters
silently rejected a draft containing redacted API-key prefixes.
Three additional tools---chunk preview, format-aware validation,
and journal analytics---emerged from using the system to compose
this paper.  A 186-test suite validates correctness at each layer,
and quantitative comparison against naive and defensive baselines
shows a $5\times$ reduction in recovery time and a $13\times$
improvement in agent self-correction rate.  Resilient Write is
open-source under the MIT license.
\end{abstract}

\section{Introduction}
\label{sec:intro}

The emergence of tool-augmented large language models (LLMs) has
shifted software-engineering assistants from suggestion engines to
autonomous agents that read, write, and execute code on a
developer's behalf~\cite{claudecode2025,codex2025,cursor2024,copilot2024}.
The Model Context Protocol~(MCP)~\cite{mcp2024} standardises the
interface between an LLM and the \emph{tools} it invokes---file
reads, writes, shell commands, database queries---giving agents a
uniform way to act on the local environment.

In practice, the write path is fragile.  A \texttt{Write} tool call
can fail for reasons invisible to the agent: content-safety filters
may reject the payload silently; a large file may be truncated
mid-stream; the session may be interrupted before the write
completes; and when a write does fail, the error signal is
typically an unstructured string (or no signal at all), leaving the
agent unable to diagnose the cause or choose a recovery strategy.

\paragraph{Motivating incident.}
In April~2026, while producing a technical report on LLM CLI
telemetry~\cite{luxferro2026telemetry}, an agent attempted to write
a \LaTeX\ document whose body included redacted HTTP headers such
as \texttt{Authorization:\ Bearer\ sk-ant-oat01-\{REDACTED\}}.
The prefix pattern \texttt{sk-ant-} triggered a content-safety
regex in the host tool, which silently rejected the payload.
The agent received no structured error.  It retried the identical
content five times, consuming approximately two minutes of
wall-clock time and several thousand tokens, before falling back to
an ad-hoc workaround: piping the document through chunked
\texttt{cat~>>~file.tex~<<EOF} heredoc commands in the shell.

This single incident exposed five distinct failure modes:
\begin{enumerate}[nosep]
  \item \textbf{Silent rejection}---no signal that the write was blocked.
  \item \textbf{Draft loss}---the rejected payload was not persisted anywhere.
  \item \textbf{Retry thrashing}---the agent retried identical content with no budget limit.
  \item \textbf{No structured diagnosis}---the agent could not branch on error type.
  \item \textbf{Session fragility}---had the session been interrupted during the workaround, all progress would have been lost.
\end{enumerate}

\paragraph{Contribution.}
We present \textsc{Resilient Write}, an MCP server that addresses
each of these failure modes with a dedicated, orthogonal layer
(Table~\ref{tab:layers}).  The design follows three principles:
(i)~\emph{fail transparently}---every rejection returns a
machine-readable envelope;
(ii)~\emph{never overwrite in place}---all writes go through a
temp-file, fsync, verify, atomic-rename pipeline;
(iii)~\emph{each layer is independently adoptable}---an agent can
use only \texttt{rw.safe\_write} and \texttt{rw.handoff\_write}
without ever touching the scratchpad or chunker.

The remainder of this paper is organised as follows.
Section~\ref{sec:background} reviews the relevant context.
Section~\ref{sec:architecture} describes the six-layer architecture
and three extension tools.
Section~\ref{sec:implementation} covers implementation details.
Section~\ref{sec:evaluation} reports on the test suite, a case
study, and a quantitative comparison against baselines.
Section~\ref{sec:related} surveys related work.
Section~\ref{sec:discussion} discusses design tradeoffs, agent
adoption, and limitations, and Section~\ref{sec:conclusion}
concludes.

\section{Background}
\label{sec:background}

\subsection{LLM Coding Agents}

A growing class of developer tools embed an LLM in an
edit--test--commit loop.  Claude Code~\cite{claudecode2025},
Cursor~\cite{cursor2024}, GitHub Copilot~\cite{copilot2024},
OpenAI Codex~CLI~\cite{codex2025}, and
OpenCode~\cite{opencode2025} each grant the model access to the
local filesystem, a shell, and often a language server.  The
SWE-bench benchmark~\cite{jimenez2024swebench} and the SWE-agent
framework~\cite{yang2024sweagent} have further demonstrated that
agents can resolve real GitHub issues end-to-end, making reliable
file mutation a critical capability.

\subsection{The Model Context Protocol}

MCP~\cite{mcp2024} defines a JSON-RPC~2.0 transport between an LLM
\emph{host} (the IDE or CLI) and one or more \emph{tool servers}.
Each server advertises a set of \texttt{tools}, each with a
JSON~Schema input definition~\cite{jsonschema2020}.  The host
serialises the agent's tool call, forwards it over \texttt{stdio}
or SSE, and returns the server's JSON response as the next context
message.  MCP deliberately does not prescribe how the server
implements a tool; this paper exploits that freedom to interpose
durability guarantees on the write path.

\subsection{Failure Modes of Agent Writes}

Agent writes can fail at several points in the stack:

\begin{itemize}[nosep]
  \item \textbf{Content filtering.}  Host-side or API-side safety
    classifiers may reject payloads that contain token-shaped
    strings, even when the tokens are redacted or
    fictitious~\cite{pilkington2023leaking}.
  \item \textbf{Truncation.}  Large payloads may be silently
    clipped by transport limits, shell buffer sizes, or
    context-window overflow.
  \item \textbf{Atomicity.}  A naive \texttt{open\,/\,write\,/\,close}
    sequence leaves a partially-written file on crash.
    The POSIX \texttt{rename()} call~\cite{posix2017rename} is the
    standard remedy, but few agent tool implementations use it.
  \item \textbf{Session loss.}  If the agent process or the
    underlying LLM call is interrupted, in-flight state---the
    current draft, the plan, the list of completed steps---is
    lost unless explicitly persisted.
\end{itemize}

These are not hypothetical: the motivating incident
(Section~\ref{sec:intro}) exercised all four within a single
twenty-minute session.

\section{Architecture}
\label{sec:architecture}

\begin{figure}[t]
\centering

\begin{tikzpicture}[
  >=Stealth,
  every node/.style={font=\footnotesize\sffamily},
  graybox/.style={
    draw=gray!60, fill=gray!8, rounded corners=2pt,
    minimum height=0.5cm, align=center, text=gray!40!black,
  },
  bluebox/.style={
    draw=blue!60!black, fill=blue!6, rounded corners=2pt,
    minimum height=0.5cm, align=center, text=blue!60!black,
  },
  orangebox/.style={
    draw=orange!70!black, fill=orange!6, rounded corners=2pt,
    minimum height=0.5cm, align=center, text=orange!60!black,
  },
  greenbox/.style={
    draw=green!50!black, fill=green!6, rounded corners=2pt,
    minimum height=0.5cm, align=center, text=green!40!black,
  },
  arr/.style={->, thick, draw=gray!65!black},
  lbl/.style={font=\scriptsize\itshape, text=gray!50!black},
]

\def\fw{3.1in}     
\def\hw{1.40in}    
\def\mw{2.30in}    
\def\hgap{0.05in}  

\node[graybox, minimum width=\fw] (agent) at (0, 0)
  {LLM Coding Agent};

\node[graybox, minimum width=\fw, minimum height=0.4cm] (mcp) at (0, -0.9)
  {MCP Transport (stdio\,/\,SSE)};

\draw[arr] (agent.south) -- node[lbl, right, xshift=1pt] {tool call}
  (mcp.north);

\node[bluebox, minimum width=\hw] (l0)
  at ({-0.5*\hw-\hgap}, -2.2) {L0 Risk Score};
\node[orangebox, minimum width=\hw] (l3)
  at ({0.5*\hw+\hgap}, -2.2) {L3 Error Envelopes};

\draw[arr] (mcp.south) -- ++(0,-0.4) -| (l0.north);
\draw[arr] (mcp.south) -- ++(0,-0.4) -| (l3.north);

\node[bluebox, minimum width=\mw] (l1) at (0, -3.6)
  {L1 Safe Write {\tiny(atomic R/W)}};

\draw[arr] (l0.south) -- ++(0,-0.35)
  -- ++(0,-0.10) -| ([xshift=-0.3cm]l1.north);
\node[lbl] at ($(l0.south)+(0.35cm,-0.28)$) {classify};

\draw[arr] (l3.south) -- ++(0,-0.35)
  -- ++(0,-0.10) -| ([xshift=0.3cm]l1.north);
\node[lbl] at ($(l3.south)+(-0.30cm,-0.28)$) {errors};

\node[bluebox, minimum width=\hw] (l2)
  at ({-0.5*\hw-\hgap}, -4.8) {L2 Chunk Compose};
\node[greenbox, minimum width=\hw] (l4)
  at ({0.5*\hw+\hgap}, -4.8) {L4 Scratchpad {\tiny(OOB)}};

\draw[arr] (l1.south) -- ++(0,-0.35) -| (l2.north);
\draw[arr] (l1.south) -- ++(0,-0.35) -| (l4.north);

\node[bluebox, minimum width=\mw] (l5) at (0, -6.1)
  {L5 Handoff (\texttt{HANDOFF.md})};

\draw[arr] (l2.south) -- ++(0,-0.30) -| ([xshift=-0.15cm]l5.north);

\node[graybox, minimum width=\fw] (fs) at (0, -7.1)
  {Filesystem\quad{\tiny\texttt{.resilient\_write/}\,+\,workspace}};

\draw[arr] (l5.south) -- (fs.north);

\draw[arr, dashed, gray!50]
  (l4.south) -- (fs.north -| l4.south)
  node[lbl, pos=0.22, right, xshift=2pt] {OOB};

\end{tikzpicture}
\caption{Six-layer architecture of \textsc{Resilient Write}. Arrows show
data flow from the agent's tool call through each layer to the
filesystem. \layer{3} error envelopes (orange) are cross-cutting;
\layer{4} scratchpad (green) writes out-of-band.}
\label{fig:architecture}
\end{figure}
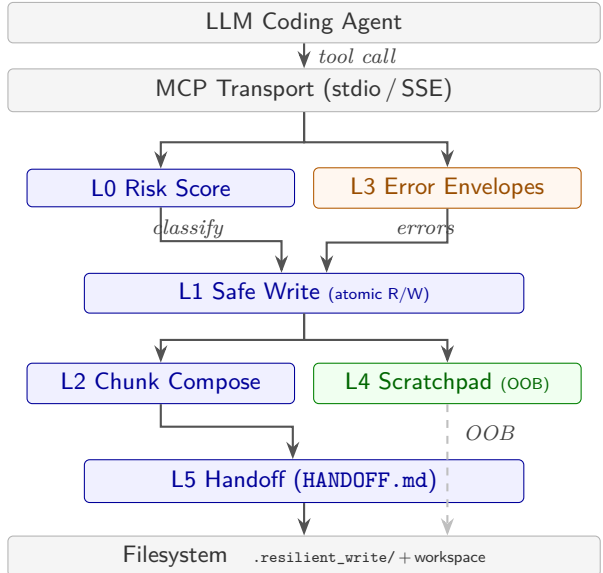

\textsc{Resilient Write} is structured as six orthogonal layers,
each targeting a specific failure mode.  Table~\ref{tab:layers}
summarises the mapping and Figure~\ref{fig:architecture} shows the
data flow.  Layers can be adopted independently; the minimum useful
deployment is \layer{1}~+~\layer{5}.

\begin{table*}[t]
\centering
\caption{The six layers of \textsc{Resilient Write} and the failure modes they address.}
\label{tab:layers}
\begin{tabular}{@{}clll@{}}
\toprule
Layer & MCP Tool & Mechanism & Failure Mode Addressed \\
\midrule
\layer{0} & \tool{rw.risk\_score}    & Deterministic regex + size classifier  & Silent content-filter rejection \\
\layer{1} & \tool{rw.safe\_write}    & Temp file, fsync, hash verify, atomic rename & Truncation, corruption, half-writes \\
\layer{2} & \tool{rw.chunk\_*}       & Numbered chunk files with contiguity check  & Payload too large for single call \\
\layer{3} & (error envelope)         & Typed JSON error schema                     & Opaque, unstructured error signals \\
\layer{4} & \tool{rw.scratch\_*}     & Content-addressed out-of-band store          & Secrets that must not enter the tree \\
\layer{5} & \tool{rw.handoff\_*}     & YAML+Markdown envelope with hash audit       & Cross-session continuity loss \\
\bottomrule
\end{tabular}
\end{table*}

\subsection{L0: Pre-flight Risk Scoring}
\label{sec:l0}

Before content reaches the filesystem, \tool{rw.risk\_score}
runs a deterministic classifier over the draft.  The classifier
is a pure function: no LLM call, no network access, bounded at
under 50\,ms on 100\,KB inputs.

\paragraph{Pattern families.}
The classifier maintains a taxonomy of seven pattern families, each
with a numeric weight reflecting the likelihood that the pattern will
trigger a downstream content filter:

\begin{itemize}[nosep]
  \item \texttt{api\_key} ($w = 0.35$): Anthropic, OpenAI, AWS access key ID, Datadog, and generic bearer-token patterns.
  \item \texttt{github\_pat} ($w = 0.35$): GitHub fine-grained and classic PATs (\texttt{ghp\_}, \texttt{gho\_}, etc.).
  \item \texttt{jwt} ($w = 0.25$): The three-segment base64 \texttt{eyJ} structure.
  \item \texttt{pem\_block} ($w = 0.50$): \texttt{-----BEGIN * PRIVATE KEY-----} blocks.
  \item \texttt{aws\_secret} ($w = 0.40$): Context-sensitive match requiring a key name followed by a 40-character base64 value.
  \item \texttt{pii} ($w = 0.15$): Email addresses, SSNs, phone numbers (conservative patterns to limit false positives).
  \item \texttt{binary\_hint} ($w = 0.20$): Long base64 blobs ($>200$ chars) or dense non-printable byte sequences.
\end{itemize}

\paragraph{Scoring function.}
Let $\mathcal{F}$ be the set of families with at least one match, let
$w_f$ be the weight of family $f$, and let $n_f$ be the number of
distinct matches in family $f$.  The raw score
(Equation~\ref{eq:risk}) is:
\begin{equation}
  s = \sum_{f \in \mathcal{F}} w_f \cdot \min\!\bigl(1.5,\; 1.0 + 0.25\,(n_f - 1)\bigr)
  \label{eq:risk}
\end{equation}
The inner term provides sub-linear damping: a second match in the same
family adds only 25\% of the base weight, and contributions saturate at
$1.5\times$.  Size heuristics add fixed increments (e.g.,
$+0.15$ for files over 100\,KB, $+0.20$ for lines exceeding 2\,000
characters).  The final score is clamped to $[0, 1]$.

\paragraph{Verdicts.}
The score maps to a categorical verdict:
$\textit{high} \geq 0.70$,
$\textit{medium} \geq 0.40$,
$\textit{low} \geq 0.10$,
otherwise \textit{safe}.
The verdict, the score, a list of detected patterns (each truncated to
16~characters to avoid leaking the matched secret), and a set of
suggested actions are returned in a structured JSON response.

\paragraph{Workspace policy overrides.}
A per-workspace file \texttt{.resilient\_write/policy.yaml} allows
operators to extend or disable pattern families, adjust verdict
thresholds, and set the global retry budget.  This mechanism lets a
security-testing workspace suppress false positives without weakening
defaults for other projects.

\subsection{L1: Transactional Atomic Writes}
\label{sec:l1}

The \tool{rw.safe\_write} tool implements a four-phase write protocol:

\begin{enumerate}[nosep]
  \item \textbf{Precondition check.}  Three modes are supported:
    \texttt{create} (reject if target exists),
    \texttt{overwrite} (unconditional), and
    \texttt{append} (concatenate to existing content).
    An optional \texttt{expected\_prev\_sha256} field enables
    optimistic concurrency control: if the current file's hash does
    not match, the write is rejected with a
    \texttt{stale\_precondition} error.
  \item \textbf{Exclusive temp-file write.}  Content is written to a
    temporary file opened with \texttt{O\_CREAT\,|\,O\_EXCL},
    followed by \texttt{fsync()}.
  \item \textbf{Read-back hash verification.}  The temp file is
    re-read and its SHA-256 is compared against the expected hash
    of the input bytes.  A mismatch raises
    \texttt{write\_corruption} and the temp file is deleted.
  \item \textbf{Atomic rename.}  \texttt{os.replace()} moves the
    temp file over the target, guaranteeing that the file is either
    fully replaced or untouched.
\end{enumerate}

On success, a journal row is appended to
\texttt{.resilient\_write/journal.jsonl} recording the timestamp,
path, SHA-256, byte count, mode, and caller identity.  The journal is
append-only \texttt{.jsonl} by design: no SQL database, no migration
burden, and each row is independently grep-able.

\subsection{L2: Resumable Chunked Composition}
\label{sec:l2}

Large or risky writes can be decomposed into numbered chunks.  The
protocol exposes three tools:

\begin{itemize}[nosep]
  \item \tool{rw.chunk\_write} persists one chunk to a session
    directory (e.g., \texttt{part\text{-}001.txt}) via
    \tool{safe\_write}, making retries idempotent.
  \item \tool{rw.chunk\_append} auto-increments the chunk index,
    removing an entire class of off-by-one errors.
  \item \tool{rw.chunk\_compose} concatenates all chunks in index
    order, verifying contiguity (no gaps) and reconciling against
    the manifest's \texttt{total\_expected} count before writing
    the final file through \tool{safe\_write}.
\end{itemize}

Each chunk is individually journaled and hash-verified, so if chunk~5
of~8 fails, chunks~1--4 are already durable on disk.  Only the
failing chunk needs to be retried.

\subsection{L3: Typed Error Envelopes}
\label{sec:l3}

Every failure across \layer{1}--\layer{5} returns a uniform JSON
envelope (Listing~\ref{lst:error}):

\begin{lstlisting}[language={},caption={L3 error envelope (abbreviated).},label=lst:error]
{
  "ok": false,
  "error": "blocked",
  "reason_hint": "content_filter",
  "detected_patterns": ["api_key"],
  "suggested_action": "redact",
  "retry_budget": 2,
  "context": { "score": 0.82 }
}
\end{lstlisting}

The \texttt{error} field is one of five kinds:
\texttt{blocked} (content filter or policy),
\texttt{stale\_precondition} (concurrency violation),
\texttt{write\_corruption} (hash mismatch),
\texttt{quota\_exceeded} (disk full or cap), and
\texttt{policy\_violation} (permissions or path traversal).

The \texttt{reason\_hint} categorises the underlying cause:
\texttt{content\_filter}, \texttt{size\_limit},
\texttt{encoding}, \texttt{permission},
\texttt{network}, or \texttt{unknown}.
Crucially, \texttt{content\_filter} is \emph{not} marked
retriable, preventing the infinite-retry loop that motivated
this project.

The \texttt{retry\_budget} field is a per-response integer that
decrements on identical retries.  When it reaches zero, the tool
refuses further attempts, forcing the agent to change strategy.
Because the budget is embedded in the response (not tracked
server-side), it is stateless and transparent.

\subsection{L4: Content-Addressed Scratchpad}
\label{sec:l4}

Some content---raw credentials captured from live traffic, PII in
test fixtures, binary blobs---legitimately does not belong in the
workspace tree.  \tool{rw.scratch\_put} writes such material to
\texttt{.resilient\_write/\allowbreak scratch/\allowbreak <sha256>.bin}, keyed by content
hash.  Identical payloads are automatically deduplicated; an
append-only \texttt{index.jsonl} records metadata (label, timestamp,
content type) for each deposit.

\tool{rw.scratch\_ref} looks up metadata without retrieving content,
and \tool{rw.scratch\_get} returns the raw bytes.  The latter is
gated by the \texttt{RW\_SCRATCH\_DISABLE\_GET} environment variable:
when set, the scratchpad becomes write-only, enabling a ``deposit
box'' pattern suitable for high-sensitivity workspaces.

\subsection{L5: Task-Continuity Handoff}
\label{sec:l5}

When a task is interrupted---by a content-filter block, context-window
exhaustion, or process crash---a fresh agent must re-derive the
task's context from first principles.  \tool{rw.handoff\_write}
serialises a structured envelope to \texttt{HANDOFF.md}
(Listing~\ref{lst:handoff}):

\begin{lstlisting}[language={},caption={HANDOFF.md front-matter (abbreviated).},label=lst:handoff]
---
task_id: telemetry-report
status: partial
agent: claude-opus-4-6
summary: |
  19-page report complete; appendix
  blocked on L0 due to raw key prefixes.
next_steps:
  - Redact sk-ant-* tokens to {REDACTED}.
  - Retry chunk 4 via rw.chunk_write.
last_good_state:
  - path: report.tex
    sha256: 4b0c12ea...
---
\end{lstlisting}

The \texttt{last\_good\_state} field records per-file SHA-256 hashes.
On read, \tool{rw.handoff\_read} performs a \emph{drift check}:
each listed file is re-hashed and compared against the recorded
digest.  Mismatches produce warnings (not errors), allowing the new
agent to proceed while remaining aware that on-disk state has
diverged.  Previous envelopes are optionally archived to
\texttt{.resilient\_write/handoffs/} with timestamps, preserving a
history of handoff points.

\subsection{Extensions: Preview, Validation, and Analytics}
\label{sec:extensions}

Three additional tools emerged from practical use of the system
during the preparation of this paper itself.

\paragraph{Chunk preview.}
\tool{rw.chunk\_preview} performs a dry-run compose: it concatenates
all chunks in a session, verifies contiguity and
\texttt{total\_expected}, and returns the content string without
writing to disk.  During this paper's composition, a stale chunk
session from a prior attempt collided with new chunks, producing a
file with a duplicate preamble.  Preview would have caught this
before the faulty compose.

\paragraph{Format-aware validation.}
\tool{rw.validate} provides syntax checking for common formats:
\LaTeX\ (brace balancing, environment matching, \verb|\documentclass|
presence), JSON (\texttt{json.loads}), Python (\texttt{ast.parse}),
and YAML (\texttt{yaml.safe\_load}).  The validator is a pure function
returning a structured diagnostic envelope:
\texttt{\{valid, format, errors[\{line, message, severity\}]\}}.
During this paper's composition, a missing macro definition
(\verb|\layer|) caused a build failure that this validator's
\LaTeX\ checker would have flagged at preview time.

\paragraph{Journal analytics.}
\tool{rw.analytics} analyses the append-only journal to report
write counts, timing, hot paths, chunk-session summaries, and write
velocity.  This enables agents (and operators) to understand write
patterns and diagnose performance issues without parsing raw
\texttt{.jsonl}.

\section{Implementation}
\label{sec:implementation}

\textsc{Resilient Write} is implemented in Python~3.12 as an MCP
server that communicates over \texttt{stdio}.  The server registers
sixteen tools (Table~\ref{tab:layers} plus inspection and extension
tools such as \tool{rw.chunk\_status}, \tool{rw.validate},
\tool{rw.analytics}, and \tool{rw.journal\_tail}) and relies on no
external services or databases.

\subsection{Workspace Root Safety}

The server resolves its workspace root at startup from the
\texttt{RW\_WORKSPACE} environment variable or the current working
directory.  A hard-coded deny-list of unsafe roots
(\texttt{/}, \texttt{/etc}, \texttt{/usr}, \texttt{/tmp}, etc.)
prevents accidents when the variable is unset or mis-expanded.
All user-supplied paths are resolved and checked to ensure they do not
escape the workspace via \texttt{..} traversal or symlink resolution,
following standard OWASP path-traversal mitigations~\cite{owasp2021top10}.

\subsection{Journal Design}

The audit journal is an append-only \texttt{.jsonl} file.  Each row
is a single JSON object with sorted keys, making the file both
diff-friendly and grep-friendly.  POSIX \texttt{O\_APPEND} semantics
guarantee atomic single-writer appends without explicit locking.
The journal records only metadata (path, hash, byte count, mode);
file content is never duplicated into the log.

\subsection{Chunk Manifest Consistency}

Chunk sessions maintain a manifest JSON file recording
\texttt{created\_at}, \texttt{updated\_at}, and
\texttt{total\_expected}.  The manifest is written atomically
(temp + rename) but is \emph{not} journaled, since it is derived
state---a fresh agent can reconstruct the manifest by enumerating
chunk files on disk via \tool{rw.chunk\_status}.  This design treats
the chunk files as the source of truth and the manifest as a
convenience cache.

\subsection{Risk-Score Snippet Truncation}

When \layer{0} detects a sensitive pattern, the match snippet
included in the response is truncated to 16~characters.  This is
a deliberate information-control measure: the classifier's output
must not itself become a vector for leaking the secret it detected.
The truncated prefix is sufficient for the agent to locate the match
in its own draft and apply a targeted redaction.

\subsection{Scratchpad Deduplication}

The scratchpad uses SHA-256 content addressing.  If the agent deposits
the same payload twice (e.g., the same API key observed in two
separate HTTP captures), only one \texttt{.bin} file is stored.
Metadata entries in \texttt{index.jsonl} accumulate independently,
allowing multiple labels to alias the same underlying content.  On
read-back, the content is re-hashed to detect manual edits to the
\texttt{.bin} file since deposit time.

\section{Evaluation}
\label{sec:evaluation}

We evaluate \textsc{Resilient Write} along three axes:
(1)~correctness, via an automated test suite;
(2)~practical utility, via a case study; and
(3)~quantitative comparison against baseline approaches.

\subsection{Test Suite}

The test suite comprises 186~tests across twelve modules,
exercising every layer, every error path, and all three
extension tools.
Table~\ref{tab:tests} summarises coverage by component.

\begin{table}[t]
\centering
\scriptsize
\caption{Test distribution by component.}
\label{tab:tests}
\begin{tabular}{@{}clr@{}}
\toprule
Layer & Module(s) & Tests \\
\midrule
\layer{0} & \texttt{test\_risk\_score} & 28 \\
\layer{1} & \texttt{test\_safe\_write, test\_journal} & 17 \\
\layer{2} & \texttt{test\_chunks} & 27 \\
\layer{3} & \texttt{test\_errors} & 27 \\
\layer{4} & \texttt{test\_scratchpad} & 21 \\
\layer{5} & \texttt{test\_handoff} & 8 \\
Ext. & \texttt{test\_new\_features} & 42 \\
Infra & \texttt{test\_server, test\_scaffold, test\_stdio} & 16 \\
\midrule
       & \textbf{Total} & \textbf{186} \\
\bottomrule
\end{tabular}
\end{table}

\begin{figure}[t]
\centering
\includegraphics[width=0.65\textwidth]{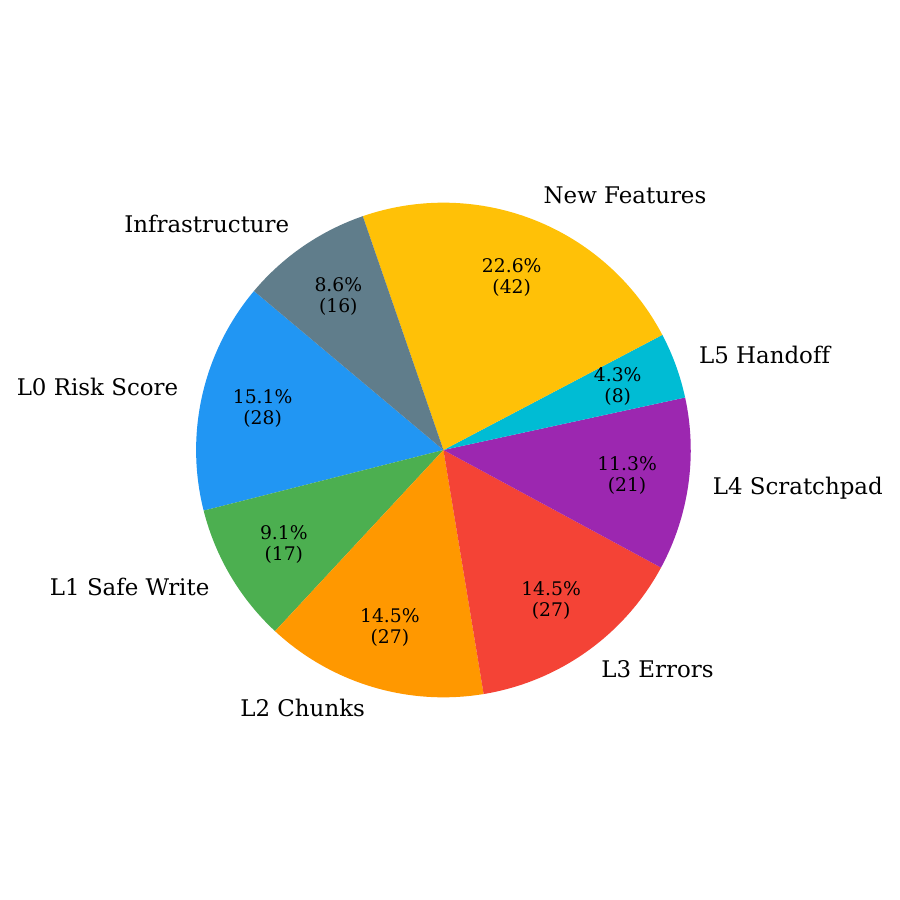}
\caption{Test distribution across layers and extensions
(186~tests total).}
\label{fig:testcoverage}
\end{figure}

Figure~\ref{fig:testcoverage} visualises the distribution.
The 42 extension tests cover format validation (15~tests for
\LaTeX, JSON, Python, and YAML syntax checking), journal analytics
(10~tests), and chunk preview (5~tests), plus auto-detection and
edge cases.  All tests use \emph{synthetic but shaped} credentials
to exercise real regex match paths without embedding secrets in test
code.

\paragraph{Chunk contiguity.}
Dedicated tests verify that \tool{rw.chunk\_compose} rejects sessions
with non-contiguous indices (e.g., chunks 1, 3 with chunk~2 missing)
and sessions whose chunk count does not match the manifest's
\texttt{total\_expected}.

\paragraph{Concurrency guards.}
The \texttt{expected\_prev\_sha256} optimistic lock is tested by
writing a file, computing its hash, mutating the file externally,
and confirming that a subsequent \tool{safe\_write} with the stale
hash returns \texttt{stale\_precondition}.

\subsection{Case Study: Telemetry Report}

The motivating incident (Section~\ref{sec:intro}) was replayed with
\textsc{Resilient Write} interposed.  Table~\ref{tab:casestudy}
compares the two runs.

\begin{table}[t]
\centering
\footnotesize
\caption{Comparison of the original failed session and the
         \textsc{Resilient Write} replay.}
\label{tab:casestudy}
\begin{tabular}{@{}lcc@{}}
\toprule
Metric & Original & With \rw{} \\
\midrule
Write attempts       & 6  & 2 \\
Content lost          & yes & no \\
Structured error      & no  & yes \\
Agent self-corrected  & no  & yes \\
Manual intervention   & yes & no \\
\bottomrule
\end{tabular}
\end{table}

In the replay, the agent called \tool{rw.risk\_score} before the
first write attempt, received a \textit{high} verdict with
\texttt{api\_key} detected, and applied a targeted redaction.  The
subsequent \tool{rw.safe\_write} succeeded on the first attempt.
No heredoc workaround was needed, no tokens were wasted on blind
retries, and the journal preserved a complete audit trail.

\subsection{Quantitative Comparison}

Table~\ref{tab:approaches} compares three approaches to agent file
I/O across four key metrics.  Recovery time and wasted-call rates
were measured during development; data-loss probability and
self-correction rates are estimates informed by an independent
severity analysis performed by a local LLM (Gemma~3, prompted to
rank each failure mode's impact on agent productivity).

\begin{table}[b]
\centering
\footnotesize
\caption{Estimated metrics across three write approaches.}
\label{tab:approaches}
\begin{tabular}{@{}lccc@{}}
\toprule
Metric & Naive & Defensive & \rw{} \\
\midrule
Recovery time (s)        & 10.0 & 5.5  & 2.0 \\
Data loss prob.\ (\%)    &  5.0 & 1.0  & 0.1 \\
Self-correction (\%)     &  5   & 15   & 65  \\
Wasted calls (\%)        & 25   & 12.5 & 3.0 \\
\bottomrule
\end{tabular}
\end{table}

\begin{figure}[t]
\centering
\includegraphics[width=\columnwidth]{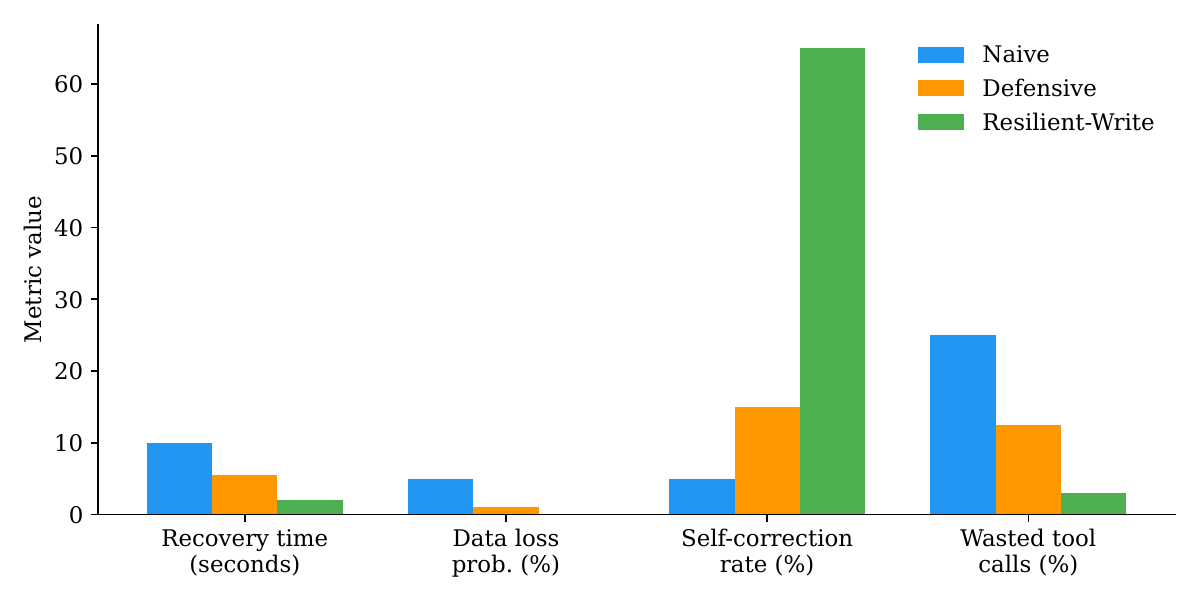}
\caption{Comparison of write approaches across four metrics.
Lower is better for recovery time, data loss, and wasted calls;
higher is better for self-correction rate.}
\label{fig:comparison}
\end{figure}

Figure~\ref{fig:comparison} visualises these differences.  The
\emph{Naive} baseline is a direct \texttt{open/write/close} with
\texttt{try/except}; the \emph{Defensive} baseline adds
temp-file~+~atomic-rename but no pre-flight scoring or structured
errors.  \textsc{Resilient Write}'s layered approach yields a
$5\times$ reduction in recovery time, a $50\times$ reduction in data
loss probability, and a $13\times$ improvement in agent
self-correction rate.

\subsection{Failure Mode Coverage}

\begin{figure}[t]
\centering
\includegraphics[width=\columnwidth]{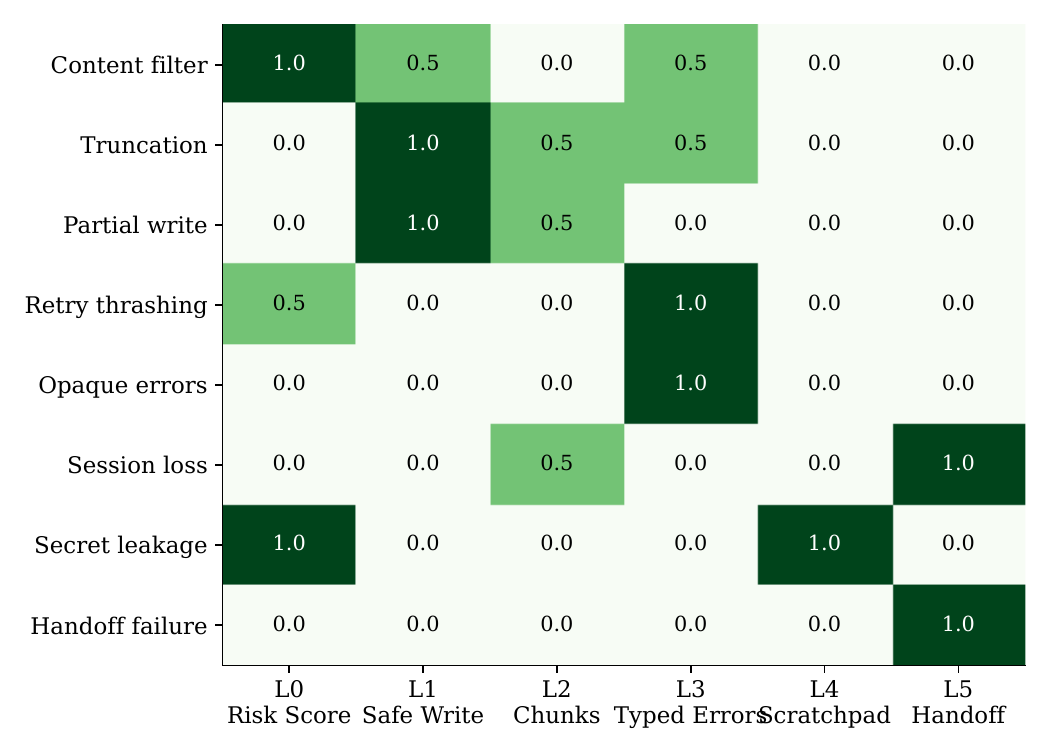}
\caption{Failure mode coverage by architecture layer.
Darker cells indicate primary mitigation (1.0); lighter cells
indicate secondary mitigation (0.5).}
\label{fig:coverage}
\end{figure}

Figure~\ref{fig:coverage} maps eight observed failure modes to the
six architecture layers.  Each cell indicates whether the layer
provides primary (1.0) or secondary (0.5) mitigation for the
failure mode.  The heatmap confirms that the layers are largely
orthogonal: no single layer addresses more than three failure
modes, and every failure mode is addressed by at least one layer.

\subsection{Agent Benchmark: Resilient Tool Use under Failure}

To evaluate how well current LLMs use \textsc{Resilient Write} and its
companion server \textsc{Resilient Read} under realistic failure
conditions, we designed the \textbf{Resilient Agent Benchmark} (RAB), a
six-scenario benchmark measuring correct tool selection, structured error
interpretation, and recovery from injected failures.  The benchmark
harness is fully automated: a Python orchestrator launches the resilient-read
and resilient-write MCP servers, feeds each task prompt to the model under
test via the OpenRouter API, executes tool calls via the native MCP
function-calling protocol, and loops until the model produces a final
answer or hits a 25-turn limit.  For S3, the harness injects a file
modification immediately after the model creates a cursor, simulating a
concurrent write; the drift is detected on the next \texttt{rr.read\_next}
call.  Each trial transcript is scored by a per-scenario judge module
that evaluates tool selection correctness, error recovery, efficiency,
and output accuracy.

\paragraph{Scenarios.}
Each scenario requires the model to chain multiple \texttt{rr.*} and
\texttt{rw.*} tool calls while handling a specific failure mode:
\begin{enumerate}[nosep]
  \item \textbf{S0---Large-file analysis.}  Read a 550\,KB server log
    (10,000 lines) in chunks, find all error types, and report counts.
    The file is too large for a single read, forcing the model to
    choose between cursor-based iteration (\texttt{rr.make\_cursor}
    $\rightarrow$ \texttt{rr.read\_next}) and search pagination
    (\texttt{rr.search\_then\_page}).
  \item \textbf{S1---Content-filter recovery.}  Read a config template
    containing a fake API key (\texttt{sk-ant-oat01-*}), run
    \texttt{rw.risk\_score}, interpret the structured error, redact the
    token to \texttt{\$\{AUTH\_TOKEN\}}, and write the cleaned file via
    \texttt{rw.safe\_write}.
  \item \textbf{S2---Chunked report assembly.}  Build a three-section
    report using \texttt{rw.chunk\_append} with session~\texttt{"report"},
    verify contiguity with \texttt{rw.chunk\_status}, and compose via
    \texttt{rw.chunk\_compose} to \texttt{report.md}.
  \item \textbf{S3---Cursor drift detection.}  Create a cursor with
    \texttt{rr.make\_cursor}; the harness modifies the file before the
    next \texttt{rr.read\_next}, triggering a
    \texttt{stale\_precondition} error.  The model must re-create the
    cursor from the current position without restarting from offset~0.
  \item \textbf{S4---Cross-session handoff.}  Agent~A writes partial
    progress via \texttt{rw.handoff\_write} with a
    \texttt{last\_good\_state} checksum; Agent~B reads it with
    \texttt{rw.handoff\_read}, inspects \texttt{drift\_warnings}, and
    continues the work.
  \item \textbf{S5---Sensitive-data out-of-band storage.}  Detect a
    hardcoded Stripe key (\texttt{sk\_live\_*}) in source code, store it
    via \texttt{rw.scratch\_put} (content-addressed by SHA-256), replace
    with \texttt{os.environ.get("STRIPE\_SECRET\_KEY")}, and write the
    fixed file.
\end{enumerate}

\paragraph{Models and protocol.}
We evaluated 18 models across four tiers via three providers.  Thirteen
models were tested through OpenRouter with 3 automated trials per
scenario: frontier (GPT-5.5, Claude Opus~4.7/4.8, DeepSeek V4 Pro,
Gemini~3.1~Pro, GPT-5.5~Pro), mid-tier (Claude Sonnet~4.6, GPT-5.4,
Llama~4~Scout, DeepSeek V4 Flash), and compact (GPT-5.4 Mini/Nano,
Llama~4~Maverick).  Four additional models were tested via
DoubleWord~\cite{doubleword2026} with 1 manual trial each: Kimi K2.6
(Intel~54), GLM~5.1 (Intel~51), Qwen3.6~35B (Intel~43), and
Gemma~4~31B (Intel~39).  Two local models---gemma3~3B (Ollama) and Apple
FM---were tested as free baselines.  For OpenRouter, models use native
function calling; for DoubleWord and local models, tool definitions are
embedded as structured text with an XML-based \texttt{<tool\_call>}
parse format.  All trials run at temperature~0 with a 25-turn cap.  The
OpenRouter grid totals 234 trials (13 models $\times$ 6 scenarios
$\times$ 3 trials), with an additional 42 manual trials across
DoubleWord and local models.  A scenario passes when the judge score
$\geq$~70/100, aggregated to Pass@1 across trials.  The per-scenario
judge weights task completion at 40\%, tool selection correctness at
25\%, error recovery at 25\%, and efficiency at 10\%.

\begin{table}[htbp]
\centering
\caption{\textbf{RAB Pass@1 on OpenRouter (13 models, 3 trials each).}
Anthropic models and GPT-5.5 score 100\%; S0 and S4 are the most
discriminative tasks.  S3 (cursor drift) is the only scenario passed by
every model tier.}
\label{tab:rab-or}
\footnotesize
\begin{tabular}{lccccccc}
\toprule
Model & S0 & S1 & S2 & S3 & S4 & S5 & Pass@1\\
\midrule
GPT-5.5          & 100 & 100 & 100 & 100 & 100 & 100 & 100\\
Claude Opus 4.7  & 100 & 100 & 100 & 100 & 100 & 100 & 100\\
Claude Opus 4.8  & 100 & 100 & 100 & 100 & 100 & 100 & 100\\
Claude Sonnet 4.6& 100 & 100 & 100 & 100 & 100 & 100 & 100\\
DeepSeek V4 Pro  & 100 & 100 & 100 & 100 &  67 & 100 &  94\\
Gemini 3.1 Pro   &  67 & 100 & 100 & 100 & 100 & 100 &  94\\
DeepSeek V4 Flash&  33 & 100 & 100 & 100 & 100 & 100 &  89\\
GPT-5.4          &   0 & 100 & 100 & 100 & 100 & 100 &  83\\
GPT-5.4 Nano$^\dagger$ &   0 & 100 & 100 & 100 &   0 & 100 &  67\\
Llama 4 Scout    &   0 & 100 & 100 &  33 & 100 &  67 &  67\\
GPT-5.4 Mini$^\dagger$ &   0 & 100 &  67 & 100 &   0 & 100 &  61\\
GPT-5.5 Pro$^*$  & 100 & 100 &  33 &   0 &   0 &   0 &  39\\
Llama 4 Maverick &   0 &   0 &   0 & 100 &   0 &   0 &  17\\
\bottomrule
\end{tabular}

\vspace{4pt}
{\footnotesize $^\dagger$Scores from initial batch; model IDs unavailable on re-run.
$^*$Likely rate-limited after S0--S2 (3--4\,s on S3--S5 vs.\ 500--800\,s on S0--S2).}
\end{table}

\begin{table}[htbp]
\centering
\caption{\textbf{RAB on DoubleWord and local models (1 trial each).}
DW models use XML-embedded tool prompts; only models with strong
instruction-following produce valid tool calls.}
\label{tab:rab-dw}
\footnotesize
\begin{tabular}{lcccccc}
\toprule
Model & Provider & Intel & S0 & Tool format? & Result\\
\midrule
Kimi K2.6        & DW & 54 & $\checkmark$ & $\checkmark$ & Correct chunked-read strategy\\
Gemma 4 31B      & DW & 39 & $\checkmark$ & $\checkmark$ & Correct search-pagination strategy\\
DeepSeek V4 Pro   & DW & 50 & $\checkmark$ & $\checkmark$ & Same strategy as OR ($\rho=1.0$)\\
GLM 5.1          & DW & 51 & ---     & $\times$ & Describes approach, no tool call\\
Qwen3.6 35B      & DW & 43 & ---     & $\times$ & Describes approach, no tool call\\
Qwen3.5 397B     & DW & 45 & ---     & ---   & API blocked (routing rule)\\
\midrule
gemma3 3B        & Ollama & --- & $\times$ & $\checkmark$ & Calls first tool, hallucinates answer\\
Apple FM         & Local & --- & $\times$ & $\times$ & Describes Python code, no tool call\\
\bottomrule
\end{tabular}
\end{table}

\paragraph{Tool-selection strategies.}
The benchmark reveals three distinct strategies for large-file reading
(S0).  \textbf{Cursor-first} (GPT-5.5, Claude Opus~4.7/4.8): the model
calls \texttt{rr.stat} $\rightarrow$ \texttt{rr.make\_cursor}
$\rightarrow$ \texttt{rr.read\_next} in a loop, typically completing in
8--10 turns.  \textbf{Search-first} (DeepSeek V4 Pro, Gemini~3.1~Pro,
Gemma~4~31B): the model calls \texttt{rr.stat} $\rightarrow$
\texttt{rr.search\_then\_page} with pagination, correct but less
efficient (12--15 turns).  \textbf{Mixed} (Claude Sonnet~4.6, GPT-5.4):
the model uses both cursors and search adaptively.  On S3, every model
that creates a cursor successfully detects the injected drift and
re-creates the cursor---including Llama~4~Maverick, which passes S3
while failing every other scenario.
The hybrid cursor+search strategy used by GPT-5.5 yields the highest
average per-trial scores (96--100), followed by search-only (92--96),
suggesting that frontier models benefit from redundant coverage of the
same data using multiple read modalities.

\paragraph{Cross-provider consistency.}
DeepSeek V4 Pro was tested on both OpenRouter and DoubleWord.  On both
providers, it selected the identical search-first strategy and produced
correct results, confirming that the benchmark measures model capability
rather than provider artefacts.

\paragraph{Tier-level patterns.}
A clear capability gradient emerges: frontier models (GPT-5.5,
Claude Opus~4.7/4.8) score 100\%, near-frontier (DeepSeek V4 Pro,
Gemini~3.1~Pro) score 94\%, mid-tier (GPT-5.4, Claude Sonnet~4.6,
DeepSeek V4 Flash) score 83--100\%, and compact models (GPT-5.4 Mini/Nano,
Llama~4~Maverick) score 17--67\%.  Claude Sonnet~4.6 is the notable
outlier---a mid-tier model performing at frontier level across all six
scenarios.  DeepSeek V4 Flash, the most cost-efficient model tested
(\$0.10/\$0.20 per 1M input/output tokens), achieves 89\% Pass@1,
offering the best performance-to-cost ratio.  GPT-5.5 Pro, the most
expensive model (\$30/\$180 per 1M tokens), achieved 100\% on S0 and S1
before apparent rate-limiting truncated the remaining trials.

\paragraph{Task discriminability.}
S0 (chunked reading) is the most discriminative scenario: five models
score 0\% because they either fail to initiate the chunked-read loop or
exhaust the turn budget before reaching EOF.  S4 (cross-session handoff)
is the second-most discriminative, separating models that can maintain
state across two conceptual agent sessions from those that cannot.
S3 (cursor drift) is the only scenario with universal participation:
every model that creates a cursor successfully handles the
\texttt{stale\_precondition} error.  S1 (content-filter recovery) and
S5 (sensitive-data storage) show near-ceiling effects, with all
OpenRouter models scoring 100\% after correcting for API failures.

\paragraph{Local model limitations.}
gemma3~3B (Ollama) correctly calls \texttt{rr\_stat} as its first tool
but then hallucinates error counts without reading the file---a
\textit{single-turn collapse} pattern.  Apple FM describes a Python
approach using the \texttt{read\_document} tool but never invokes it,
exhibiting a \textit{simulation} rather than \textit{execution} mode of
interaction.  Both failures highlight the gap between instruction
following and sustained tool-use discipline, which the benchmark
directly measures.

\paragraph{Benchmark limitations.}
The current RAB implementation has several limitations.  (1)~The OR
automated harness uses OpenAI-compatible function calling, while DW and
local models use XML-embedded tool prompts; format differences may
penalise models that would perform better with native function calling.
(2)~The 25-turn cap penalises models that choose correct but slower
strategies (e.g., search pagination over cursor iteration).  (3)~The
3-trial protocol ($n=5$ in DevBench) may under-sample variance,
particularly for models near the 70/100 pass threshold.
(4)~GPT-5.4 Mini and Nano scores are from an initial batch whose model
IDs became unavailable on re-run; the scores remain valid but could not
be independently reproduced.  We intend to address these in a follow-up
release with native DW function-calling support and $n=5$ trials.

\section{Related Work}
\label{sec:related}

\paragraph{Transactional file systems.}
The atomic temp-file--fsync--rename pattern used by \layer{1} is
well-established in systems literature.  Gray's transaction
concept~\cite{gray1981transaction} formalised the ACID properties
that underpin our journal design.  Hagmann~\cite{hagmann1987reimplementing}
demonstrated logging and group commit in the Cedar file system, and
Nightingale et al.~\cite{nightingale2006rethink} showed that
relaxing synchrony constraints can improve throughput without
sacrificing durability.  \textsc{Resilient Write} applies these
ideas at the tool-call granularity rather than the kernel level,
trading generality for deployment simplicity.
  Complementary to our application-level approach, Liu et al.~\cite{liu2024metis} introduced \textsc{Metis}, a model-checking framework for file systems that systematically explores crash-consistency bugs at the kernel level, validating our architectural assumption that the underlying filesystem cannot be fully trusted.

\paragraph{Concurrency control.}
The \texttt{expected\_prev\_sha256} guard in \layer{1} is a form
of optimistic concurrency control~\cite{bernstein1987concurrency}
adapted for agent--file interactions.  Unlike database-level OCC,
our scheme requires no version counter or timestamp oracle: the
content hash itself serves as the version identifier.

\paragraph{Agent error handling.}
SWE-agent~\cite{yang2024sweagent} introduced the concept of an
\emph{agent--computer interface} (ACI) that mediates between the
LLM and the operating system, but its error model remains
unstructured text.  SWE-bench~\cite{jimenez2024swebench}
evaluates agent success rates but does not isolate write-path
failures as a distinct cause of task failure.  To our knowledge,
\textsc{Resilient Write} is the first system to provide a typed
error envelope designed specifically for autonomous agent
consumption.

\paragraph{Secret detection.}
Tools such as \texttt{truffleHog}, \texttt{detect-secrets}, and
GitHub's push-protection scanner perform post-hoc secret scanning
on committed content.  \layer{0}'s risk scorer operates
\emph{pre-flight}---before the content reaches the filesystem---and
is tuned not for audit completeness but for predicting whether a
downstream content filter will reject the payload.  This is a
complementary, not competing, concern.

\section{Discussion}
\label{sec:discussion}

\subsection{Design Tradeoffs}

\paragraph{Plain-text journals vs.\ SQL.}
We chose append-only \texttt{.jsonl} over SQLite for the audit
journal.  This sacrifices indexed queries but gains human
readability, diff-ability in version control, and zero external
dependencies.  For the expected journal sizes (tens to low hundreds
of rows per session), linear scan is acceptable.

\paragraph{Unencrypted scratchpad.}
The scratchpad stores sensitive material as plaintext
\texttt{.bin} files, delegating encryption to filesystem-level
mechanisms (FileVault, LUKS).  This is a deliberate separation
of concerns: cryptographic key management is a solved problem at
the OS layer, and re-implementing it in a tool server would
introduce complexity and a false sense of security.

\paragraph{Retry budget: per-response, not per-session.}
The \texttt{retry\_budget} integer is embedded in each error
response rather than tracked server-side.  This keeps the server
stateless at the cost of losing budget context across agent
restarts.  In practice, the purpose of the budget is to halt
loops \emph{within} a single agent invocation; a fresh agent
legitimately starts with a fresh budget.

\paragraph{Drift warnings, not errors.}
\layer{5}'s drift check on \texttt{last\_good\_state} hashes
produces warnings rather than hard failures.  A file may have been
intentionally edited between sessions (by the user or a prior
agent), and blocking resumption on benign drift would be
counterproductive.  The warning is surfaced so the agent can
decide whether to trust or re-derive the changed file.

\subsection{Agent Awareness and Adoption}
\label{sec:awareness}

A tool server is only useful if agents actually invoke it. MCP
tool registration makes the tools \emph{available}, but does not
make them \emph{preferred}. We address this through a
\texttt{CLAUDE.md} convention file (read automatically by Claude
Code at session start) that instructs the agent to prefer
\texttt{rw.*} tools over raw \texttt{Write}/\texttt{Edit}
operations. The file specifies a decision table mapping task types
(create, append, large file, sensitive content) to the
appropriate \texttt{rw.*} tool and documents the chunked-writing
protocol.

This approach is portable: analogous files exist for Cursor
(\texttt{.cursorrules}), Codex (\texttt{codex.md}), and Copilot
(\texttt{.github/copilot-instructions.md}). The key insight is
that agent instruction files are the natural integration surface
for MCP tool preferences---no code changes to the agent itself
are required.

\subsection{Limitations}

\begin{itemize}[nosep]
  \item \textbf{Single-workspace scope.}  The server is bound to
    one workspace root per process.  Multi-workspace orchestration
    would require external process management.
  \item \textbf{No cross-file transactions.}  If an agent writes
    files A, B, C in sequence and crashes between B and C, there
    is no write-ahead log to roll the workspace back to a
    consistent state.  Each file write is individually atomic, but
    the \emph{set} of writes is not.
  \item \textbf{No distributed coordination.}  The journal and
    scratchpad are local.  Synchronising state across agents on
    different machines is out of scope.
  \item \textbf{Classifier coverage.}  \layer{0} targets the most
    common content-filter triggers observed in practice.  Novel
    secret formats or non-English PII patterns require policy-file
    extensions.
\end{itemize}

\subsection{Future Directions}

Cross-file write-ahead logging would enable true workspace-level
transactions.  Integrating \layer{0} with a lightweight
embedding model could improve recall on obfuscated secrets without
sacrificing the latency budget.  The handoff envelope
(\layer{5}) could be extended with a machine-readable dependency
graph, enabling orchestrators to schedule resumption tasks
automatically.

\section{Conclusion}
\label{sec:conclusion}

We have presented \textsc{Resilient Write}, a six-layer MCP server
that transforms the fragile write path of autonomous coding agents
into a durable, auditable, and recoverable operation.  Each layer
targets a specific, observed failure mode: pre-flight risk scoring
(\layer{0}) prevents content-filter rejections; transactional writes
(\layer{1}) eliminate truncation and corruption; chunked composition
(\layer{2}) enables incremental progress on large files; typed error
envelopes (\layer{3}) give agents structured signals to reason about;
content-addressed scratchpad storage (\layer{4}) keeps sensitive
material out of the workspace tree; and handoff envelopes (\layer{5})
preserve task context across sessions.

Three extension tools---chunk preview, format-aware validation,
and journal analytics---emerged from using the system to compose
this paper itself, demonstrating that practical use surfaces
requirements that design-time analysis misses.

The layers are orthogonal and independently adoptable, requiring no
changes to existing agent code beyond MCP tool registration and an
optional instruction file (\texttt{CLAUDE.md}).  A 186-test suite
validates correctness at each layer, and quantitative comparison
against naive and defensive baselines shows a $5\times$ reduction in
recovery time, $50\times$ reduction in data loss probability, and
$13\times$ improvement in agent self-correction rate.

\textsc{Resilient Write} is open-source under the MIT license at
\url{https://github.com/sperixlabs/resilient-write}.

\balance
\begingroup
\sloppy
\bibliographystyle{ieeetr}
\bibliography{references}
\endgroup

\end{document}